\documentclass{llncs}
\usepackage{hyperref}
\usepackage{graphicx}
\begin{document}

\title{Senior Programmers: Characteristics of Elderly Users from Stack Overflow}
\titlerunning{Elderly users on Stack Overflow}  
%
\author{Grzegorz Kowalik \and Radoslaw Nielek}
\authorrunning{Grzegorz Kowalik et al.} 
%
\tocauthor{Grzegorz Kowalik and Radoslaw Nielek}
\institute{Polish-Japanese Academy\\
of Information Technology\\
Koszykowa 86\\
02-008 Warsaw, Poland\\
\email{(grzegorz.kowalik, nielek)@pjwstk.edu.pl}}

\maketitle              

\begin{abstract}
In this paper we presents results of research about elderly users of Stack Overflow (Question and Answer portal for programmers). They have different roles, different main activities and different habits. They are an important part of the community, as they tend to have higher reputation and they like to share their knowledge. This is a great example of possible way of keeping elderly people active and helpful for society.  
\keywords{elderly, Stack Overflow, Q\&A, aging, online communities, programmers, crowdsourcing}
\end{abstract}

\section{Introduction}

When we think about programmers, we usually imagine relatively young people sitting in front of hi-tec computers. They learn fast and are able to follow rapid changes in technology and new trends. But do they get older? Some of them become senior programmers but this phrase is related to the experience and not age. On the other hand computers and programming languages have been with us already for over sixty years, thus, there should exist also somewhere people who operated computer systems in sixties and seventies. What are they doing now. Do they learn new skills and languages?

In this paper we take a closer look on elderly users of Stack Overflow\footnote{http://stackoverflow.com} -- the biggest online community for programmers according to Alexa rating\footnote{http://www.alexa.com/siteinfo/stackoverflow.com} with over 5.7 million registered users and 31 million posts.

The very first question that appears is whether older adults use the Stack Overflow at all. Galit Nimrod \cite{nimrod2010seniors} identified 40 online communities where older adults discuss health, tourism or retirements related issues, so there is no reason to assume that technical savvy seniors avoid Q\&A web sites devoted to programming but what roles do they play in the community? Do they overwhelmingly look for information or maybe they are experts in long-forgotten technologies and programming languages? How different their habits are in comparison with younger users?

Three things make Stack Overflow a particularly interesting place for studying older adults behaviors. First, there are users of very different age range, starting from thirteen or fourteen years of age, thus a whole spectrum of inter-generational behaviors might emerge. Second, it is a community purely oriented on knowledge exchange in a fast pacing topics, so being active requires either an in-depth knowledge or readiness to learn. Finally, all data are openly available and coupled with meta-information such as up and downvotes, timestampes and tags for posts and reputation for users.

Understanding how older adults can benefit from online communities and how they can contribute to the communities' goals is crucial for long-term development of online Q\&A sites in the light of dramatically increasing proportion of elderly in society. Active participation in Q\&A sites is also a sign of long-life learning which is important to keep people in labor market. 

The remaining part of the paper is organized as follow: next section presents related work, dataset used in this paper is described in section \ref{data_source}. Data analyses and results are presented in section \ref{sec:results}. Possible further studies and conclusion are gathered in section \ref{sec:conclusion}.

\section{Related work}

The aging and programming skills was researched before using Stack Overflow data in \cite{morrison_is_2013}. Their research shows the differences between older and younger users, and prove that elderly users have better reputation. They also found that elderly users are still learning new technologies.

The concept of deeper differences between older and younger groups is a well known issue for sociologists. There are even studies dedicated to the generations related to technology \cite{mcmullin_generational_2007}, \cite{robat_history_2006}. Results show that we can describe IT workers from some technological inventions as "generations" of some technology and they also find themselves a part of it.

About activation and using knowledge and skills of elderly people in crowd sourcing, there were experiments and research showing that they might not be interested in doing small tasks on demand in exchange for small remuneration, although they are interested in "being useful" for society, like proofreading \cite{itoko2014involving,kobayashi2013age,kobayashi2015motivating} or Amazon mTurk small tasks \cite{brewer2016would}. Some works were focused on similar topic as this paper - knowledge sharing \cite{hiyama2013question}. 

Stack Overflow is also an example of an online community. Regarding elderly people in such communities, there were some research about them like: \cite{nimrod2013probing} where authors show different types of users ( information swappers, aging-oriented, socializers), roles of online communities for elderly \cite{berdychevsky2015let}, showing that they can give them both entertainment and valuable information, or \cite{nimrod2010seniors}, where authors show Natural Language Processing method of measuring users engagement. Research in \cite{nimrod2014benefits} shows that being a part of an online community can enhance elderly people well-being. Topic of positive influence of IT technologies on elderly people well being is also researched in \cite{dickinson2006computer}. Other topics of discussion than programming were researched in \cite{nimrod2012online} (tourism), where we can also see knowledge and information sharing, similar to Stack Overflow.

\section{Data sources}
\label{data_source}

All results presented in this paper are based on the analysis of one of two data sources (sometimes on both): 
\begin{itemize}
\item Stack Exchange API\footnote{\url{https://data.stackexchange.com/}} -- tool shared by an administrator that makes possible to submit SQL-like queries and receives results directly from Stack Overflow database; all elements existing in the web site can be gathered that way, i.a.: reputation, users public profiles, posts, topics, tags; the only limitation is a cap of 50 thousands rows per query, thus in some cases we had to do calculation on a sample (always clearly stated in text if applicable);
\item Stack Overflow Survey -- a survey of Stack Overflow users is conducted every year; the most recent results available are from year 2015\footnote{\url{http://stackoverflow.com/research/developer-survey-2015}}; among 25831 users who answered questionnaire only $0.5\%$ were 60+, therefore in some cases it was impossible to calculate the significance tests or verify hypothesis requiring intersection of many variables.
\end{itemize}

\section{Results}
\label{sec:results}
\subsection{Do older adults use Stack Overflow?}

\begin{figure}
\centering 
\includegraphics[width=110mm]{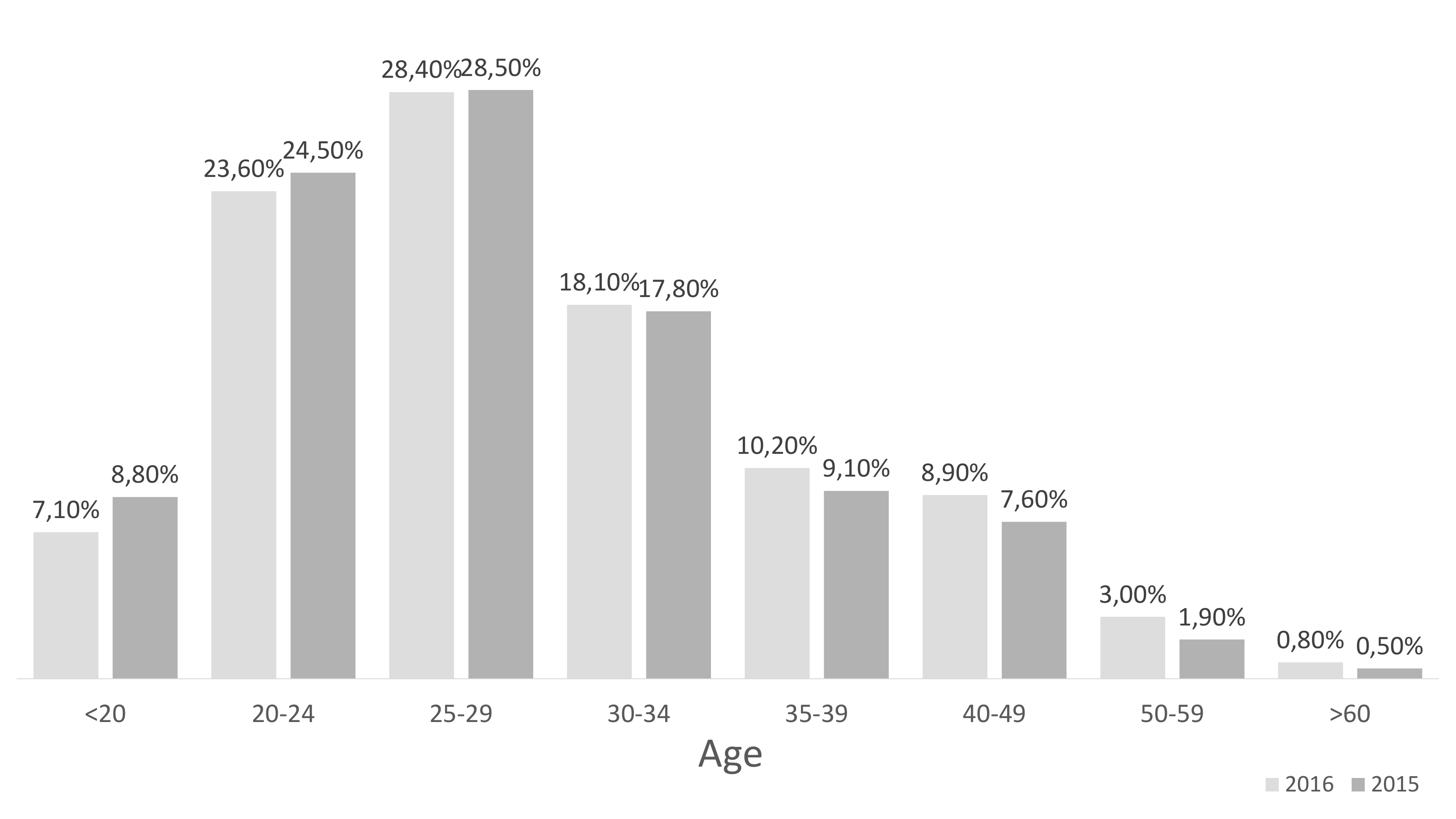}
\caption{Age (survey data). N=55338 (2016) and N=25831 (2015)}
\label{age_survey}
\end{figure}

As "seniors" (elderly people) we call users that are 60 or more years old. This is mainly because of the limitations of survey data where age was given in range. Figure \ref{age_survey} shows the distribution of age for surveys conducted in 2015 and 2016\footnote{In the rest of the paper we use data from 2015 instead of 2016 as not all data are still available for the newest survey.}. Bins reflect possible answers in this question. There is 0.5\% -- 0.8\% of users with age 60+.

\begin{figure}[!ht]
\centering 
\includegraphics[width=120mm]{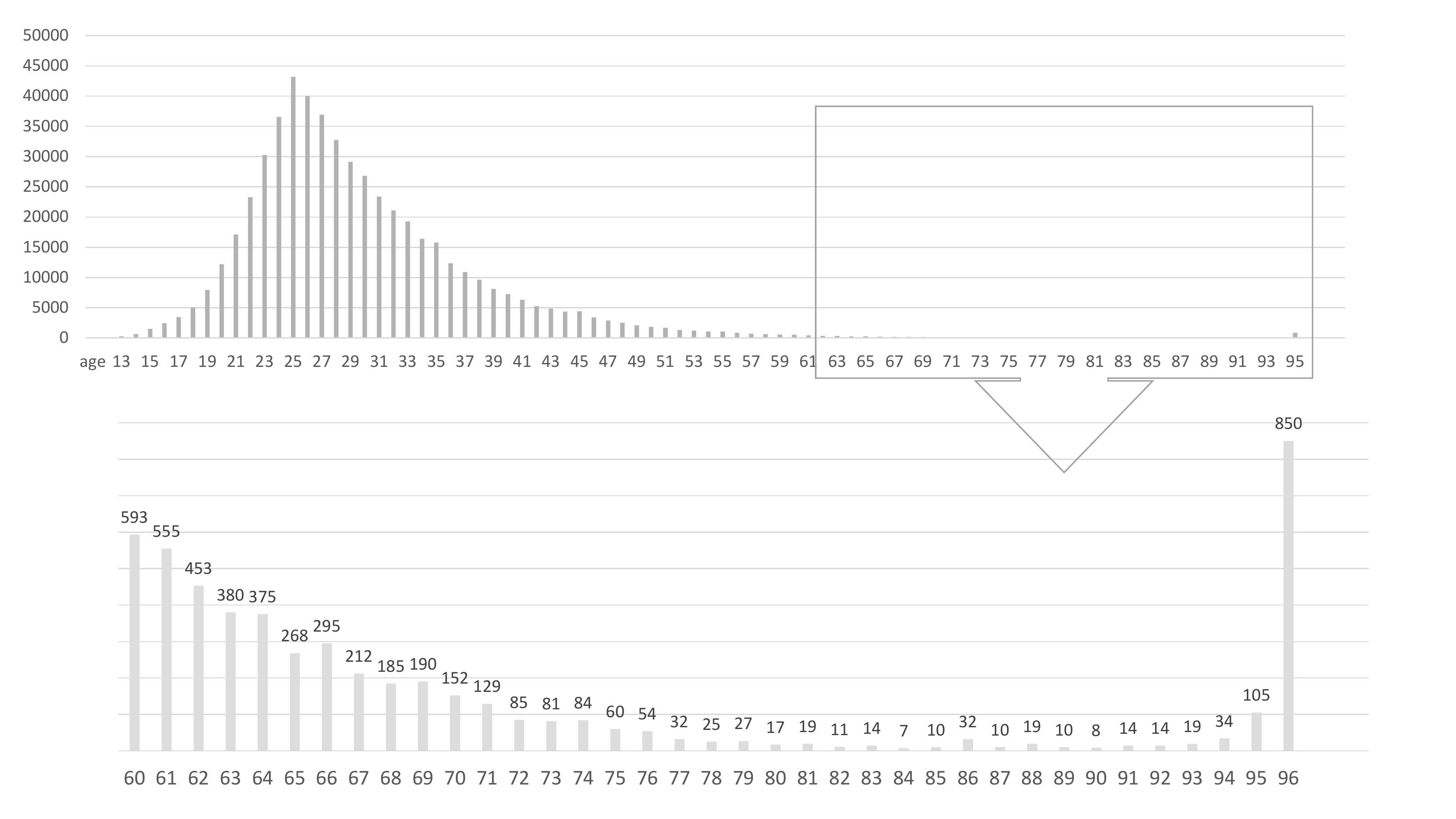}
\caption{Age distribution (number of users) among Stack Overflow users}
\label{age}
\end{figure}

Next to the survey age can also be extracted from users profiles but we have to be aware that providing birth year is not mandatory and users may also lie. We therefore excluded all users without birth year in their profile, assuming that providing age is not correlated with any important variables. Moreover, we have taken a closer look at age distribution from Stack Overflow users profile data. Age distribution from both sources look similar with one exception. Closer look at Figure \ref{age} reveals surprisingly many users over that are over 90. We should expect age to have a more Gaussian shape, thus we can assume that these profiles are fake. We decided to exclude profiles with age over 90.

After excluding users with "too high" age, we received 562795 users (0.8\% of them have a age 60 or more). This is very similar to the survey results (Figure \ref{age_survey}) in 2016, so we can assume that our filtering was justified. To further confirm that there are real older adults in on Stack Overflow we investigated users self-descriptions published on users profile web pages. In-depth insight can be found in section \ref{sec:profile_texts}. 


\subsection{Do they teach or do they learn?}

If we assume that older adults have more experience and are more knowledgeable, we should observe that they are posting overwhelmingly more answers instead of questions. How much effort do they put in developing the community? We try to answer these questions by comparing frequency of post types, length of profile texts and responses about motivation (taken from survey) between two age groups.

\subsubsection{Post types: main activity}

In Stack Overflow users can be active in several ways: writing posts, comments, upvoting, downvoting etc. The most important user activity is posting. Posts can have different types: question or answer. Asking questions we can interpret as learning -- looking for knowledge, and by answering -- teaching, sharing knowledge. Our hypothesis is, that elderly people are less likely to learn, and are more interested in sharing their knowledge -- comparing to younger users.

To verify this, we used post type table and frequency of different post types in each age group. As you can see in Table \ref{post_type}, answers are the most popular post type in both groups. This is probably due to the fact that for each question you can have more than one answer, so it is not surprising that both groups have more answers than questions in their posting activity. Moreover, instead of posting a new question users can always read answers for similar questions that are already posted and answered.

\begin{table}
\centering
\caption{Post types frequency in each age group. Remain four types of posts has been omitted because are extremely rare in Stack Overflow -- less than 1\%.}
\label{post_type}
\begin{tabular}{|c|c|c|}
\hline
\textbf{PostTypeId}          & \textbf{Juniors} & \textbf{Seniors} \\ \hline
\textbf{Question}            & 25.04\%          & 13.54\%          \\ \hline
\textbf{Answer}              & 74.61\%          & 86.26\%          \\ \hline
\end{tabular}
\end{table}

It is the difference that is important here -- "Seniors" have significantly higher percentage of answers in their posts than "juniors" (Significance t-test for difference of answers ratio in both groups have p-value bellow 0.0001). This confirms our hypothesis -- "seniors"  more often (more than 10 percentage points) gives answers, share their knowledge, than ask questions. They more teach than learn.

\subsubsection{Profile texts: effort}

We also expect elderly people to be more engaged in Stack Overflow community. This is already somehow confirmed in the previous point, as giving more answers is also creating a content for portal and being responsible for it. In addition, we checked their profile texts. We assume that if they put more effort in writing about themselves, they care more about the community.

As we can see in Table \ref{profile_text_no}, "Seniors" have significantly (verified by T-Test), higher average of characters used in their profile texts -- almost two times higher. This confirms our hypothesis that they put more effort into their activity.

\begin{table}
\centering
\caption{Profile text length (number of characters)}
\label{profile_text_no}
\begin{tabular}{|c|c|c|ll}
\cline{1-3}
                                                             & \textbf{Seniors} & \textbf{Juniors} &  &  \\ \cline{1-3}
\textbf{Mean}                                                & 180              & 92               &  &  \\ \cline{1-3}
\textbf{\begin{tabular}[c]{@{}c@{}}Standard\\   Deviation\end{tabular}} & 342.22           & 209.40           &  &  \\ \cline{1-3}
\textbf{N}                                               & 4273             & 528847           &  &  \\ \cline{1-3}
\end{tabular}
\end{table}

\subsubsection{Motivation}

\begin{figure}
\centering 
\includegraphics[width=130mm]{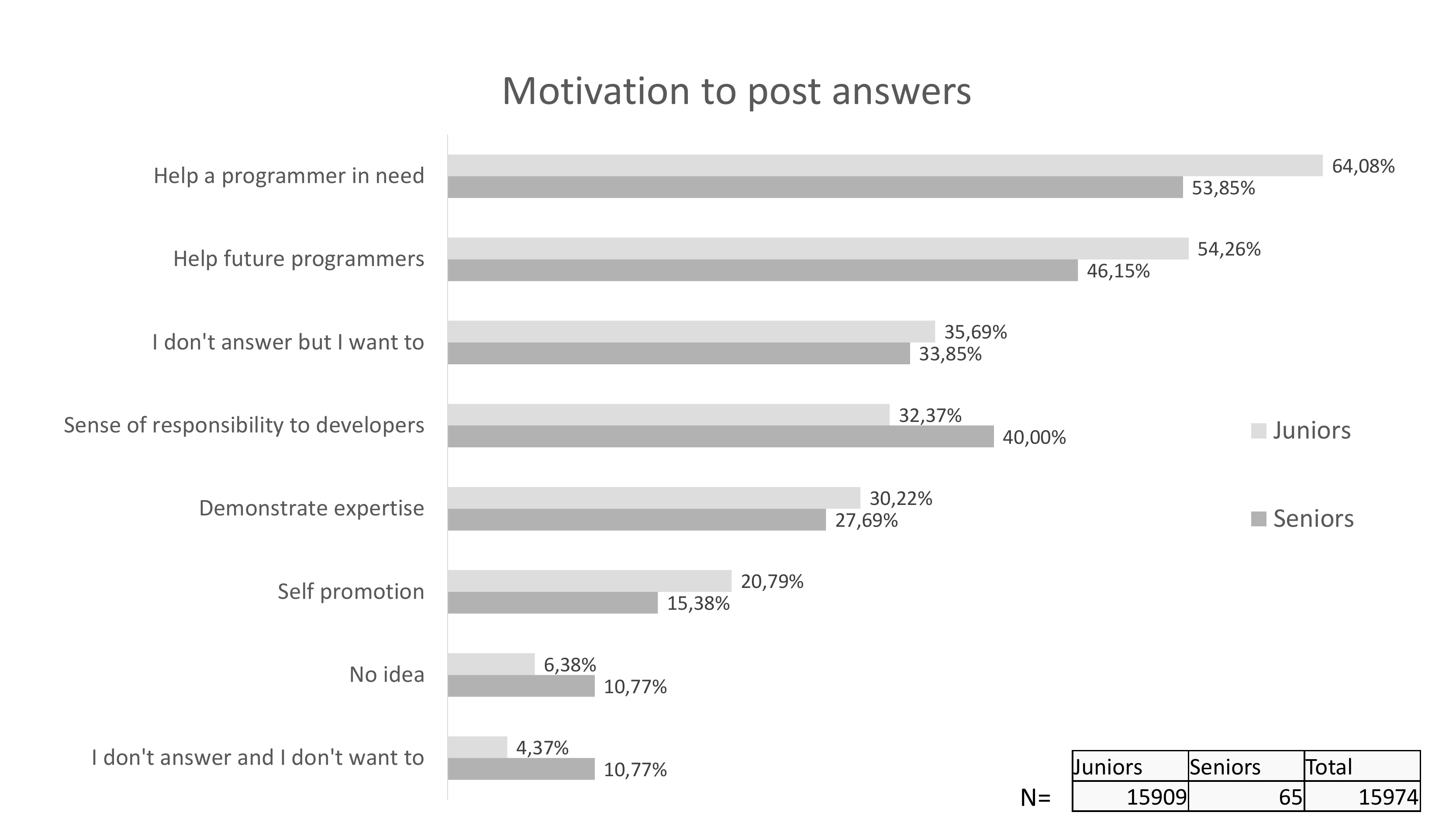}
\caption{Motivation to post answers (survey data)}
\label{motivation}
\end{figure}

In survey, there was a question regarding motivation of posting answers to questions in Stack Overflow. Results, divided into our age groups, are presented in Figure \ref{motivation}. We can learn that "helping programmers in need" and "future programmers" are most important for both groups. We also observe the differences in motivation with older adults biased more toward sense of responsibility but because of small sample differences are not statistically significant (sample of seniors that given answer is too small).

\subsection{Earned reputation or gamed reputation?}

Older adults share their knowledge by posting more answers than younger users but are their answers valuable for the community? In order to check it, we decided to take a closer look on the reputation earned by two group of users. According to Stack Overflow \textit{"Reputation is a rough measurement of how much the community trusts you; it is earned by convincing your peers that you know what you’re talking about."}\footnote{\url{http://stackoverflow.com/help/whats-reputation}}. As we can see in Table \ref{rep}, older adults have higher reputation (significance test confirms it). "Seniors" have higher mean of reputation, but also, as we could see also in paper \cite{morrison_is_2013}, much higher standard deviation.

\begin{table}
\centering
\caption{Reputation overview}
\label{rep}
\begin{tabular}{|c|c|c|}
\hline
\textbf{Age}     & \textbf{Mean} & \textbf{St. Dev} \\ \hline
\textbf{Juniors} & 645           & 5630.17          \\ \hline
\textbf{Seniors} & 938           & 11165.99         \\ \hline
\end{tabular}
\end{table}

Reputation have very high standard deviation in general - as it is not mandatory to post, a lot of users have low activity and low reputation. Only around a half have the reputation higher than 1. This way we need to look closer on the results than mean.

\begin{table}
\centering
\caption{Reputation percentiles (for "Juniors" we used sampling (around 20\% of total))}
\label{rep_perc}
\begin{tabular}{|c|c|c|}
\hline
\textbf{Percentile} & \textbf{Juniors*} & \textbf{Seniors} \\ \hline
\textbf{20}          & 1               & 1               \\ \hline
\textbf{40}        & 5               & 1               \\ \hline
\textbf{60}        & 37              & 11              \\ \hline
\textbf{80}        & 276             & 101             \\ \hline
\textbf{90}        & 833.9           & 456             \\ \hline
\textbf{100}          & 265086          & 446919          \\ \hline
\end{tabular}
\end{table}

As we can see in Table \ref{rep_perc}, reputation is distributed in different ways among both groups. Among "Juniors", there are more active users, around 40\% of them have reputation higher than 1. For "seniors" its around 50\%. After this point, we can see the reputation distributed in a more equal way among "juniors" and less equal in "seniors" group. That also match the higher variation in "seniors" group.

In addition, we also checked how often answers of "seniors" and "juniors" were marked as "accepted" (best answers for question). "Seniors" have 32.98\% answers marked as accepted, "juniors" have  36.04\% .  As we can see juniors have higher percentage of accepted answers, but this is not a high difference. This might have many reasons, as accepted answers are usually those provided fast or younger users have some better strategies \cite{adamska_study_2015}. 

\subsection{Profile texts: Recognizing themselves as elderly}
\label{sec:profile_texts}
As we were processing profile texts, we noticed that elderly people describe themselves using age-related words. Below there are few examples from profiles of older adult users with reputation bigger than 1: 
\begin{itemize}
\item "\textit{I'm an old guy who likes programming, photography, chess, science fiction, and music.}",
\item "\textit{Old-ish IT Geezer, young at heart, memoir fanboy}",
\item "\textit{retired, learning python to help save what grey cells are left.}".
\item \textit{"Started programming in 1980 at SAC headquarters in Omaha, NE. I worked on 3D applications for B-52 and Cruise Missile mission planning. (...)"},
\item \textit{"Software Architect, aspiring writer. Programmer for well over 30 years, about 70\% of my working life."},
\item \textit{"Grandfather, programmer, vegan."}
\end{itemize}

We clearly see, that their age in profiles is not fake, as they identify with old age in their profile texts. They use words like "old", "ancient", "grandfather", etc. 


\subsubsection{Survey: Different habits - IDE}

We were also looking for different habits related to programming among elderly users. There is significant difference in preferred IDE (Integrated Development Environment), as we can see in Table \ref{ide}, elderly users prefer light IDE (light/white background with dark/black fonts), and younger prefers opposite.

\begin{table}
\centering
\caption{Preferred IDE (survey data)}
\label{ide}
\begin{tabular}{|c|c|c|}
\hline
                       & \textbf{Seniors} & \textbf{Juniors} \\ \hline
\textbf{Dark}  	& 9.23\%            & 52.65\%          \\ \hline
\textbf{Light} & 64.62\%           & 7.61\%          \\ \hline
\textbf{don't use IDE}    & 26.15\%          & 39.76\%          \\ \hline
\end{tabular}
\end{table}

This might be an important information for tool developers, but also significant example of differences between younger and elder programmers. This can be explained by medical aspects, considering sight problems, but also with their habits. There is a consensus saying that for elderly people there should be a high contrast in the interface and the best is the combination of black and white. However, if the background should be black or white and font opposite, there is less consent.

Some papers show that black background is better, like in \cite{slavicek_designing_2014}. Other say opposite like \cite{carmien_elders_2014}. White background is also recommended in medical articles, like in \cite{bauer_improving_1980}. However, we must remember that technology is changing rapidly, and a lot depends on screen type etc. In some cases this might be also a lack of ability to change IDE, but it is not likely for such group like programmers. We are not sure if they prefer it because of old habits or medical issues. 

\section{Conclusions}
\label{sec:conclusion}

To sum up, we can say that elderly users constitutes an important part of the community and they feel more engaged in it. We had no possibility of measuring factors like well-being, but we can expect, according to similar cases in \cite{nimrod2014benefits} and \cite{dickinson2006computer}, that it positively influences it. They are recognized and awarded by reputation, their posts are up voted, they feel rewarded and important - and in fact, they are, because with their answers they build a portal (crowd sourcing way) that is used worldwide. Everyday programmers looks for answers in Stack Overflow and use their knowledge. 

As computers, mobile technology and internet are becoming more and more popular, the next generations of older adults will be more willing to be a part of such online communities. It might be a great way of keeping the elderly people active.

\subsubsection*{Acknowledgements.}This project has received funding from the European Union’s Horizon 2020 research and innovation programme under the Marie Sklodowska-Curie grant agreement No 690962

\bibliography{Zotero,ref_dream}{}

\begin{thebibliography}{10}

\bibitem{adamska_study_2015}
Paulina Adamska and Marta Juźwin.
\newblock Study of the {Temporal}-{Statistics}-{Based} {Reputation} {Models}
  for {Q}\&{A} {Systems}.
\newblock {\em Computer Science}, 16(3):253, September 2015.

\bibitem{bauer_improving_1980}
D.~Bauer and C.~R. Cavonius.
\newblock Improving the legibility of visual display units through contrast
  reversal.
\newblock {\em Ergonomic aspects of visual display terminals}, pages 137--142,
  1980.

\bibitem{berdychevsky2015let}
Liza Berdychevsky and Galit Nimrod.
\newblock " let's talk about sex": Discussions in seniors' online communities.
\newblock {\em Journal of Leisure Research}, 47(4):467, 2015.

\bibitem{brewer2016would}
Robin Brewer, Meredith~Ringel Morris, and Anne~Marie Piper.
\newblock Why would anybody do this?: Understanding older adults' motivations
  and challenges in crowd work.
\newblock In {\em Proceedings of the 2016 CHI Conference on Human Factors in
  Computing Systems}, pages 2246--2257. ACM, 2016.

\bibitem{carmien_elders_2014}
Stefan Carmien and Ainara~Garzo Manzanares.
\newblock Elders {Using} {Smartphones} – {A} {Set} of {Research} {Based}
  {Heuristic} {Guidelines} for {Designers}.
\newblock In Constantine Stephanidis and Margherita Antona, editors, {\em
  Universal {Access} in {Human}-{Computer} {Interaction}. {Universal} {Access}
  to {Information} and {Knowledge}}, number 8514 in Lecture {Notes} in
  {Computer} {Science}, pages 26--37. Springer International Publishing, June
  2014.
\newblock DOI: 10.1007/978-3-319-07440-5\_3.

\bibitem{dickinson2006computer}
Anna Dickinson and Peter Gregor.
\newblock Computer use has no demonstrated impact on the well-being of older
  adults.
\newblock {\em International Journal of Human-Computer Studies},
  64(8):744--753, 2006.

\bibitem{hiyama2013question}
Atsushi Hiyama, Yuki Nagai, Michitaka Hirose, Masatomo Kobayashi, and Hironobu
  Takagi.
\newblock Question first: Passive interaction model for gathering experience
  and knowledge from the elderly.
\newblock In {\em Pervasive Computing and Communications Workshops (PERCOM
  Workshops), 2013 IEEE International Conference on}, pages 151--156. IEEE,
  2013.

\bibitem{itoko2014involving}
Toshinari Itoko, Shoma Arita, Masatomo Kobayashi, and Hironobu Takagi.
\newblock Involving senior workers in crowdsourced proofreading.
\newblock In {\em International Conference on Universal Access in
  Human-Computer Interaction}, pages 106--117. Springer, 2014.

\bibitem{kobayashi2015motivating}
Masatomo Kobayashi, Shoma Arita, Toshinari Itoko, Shin Saito, and Hironobu
  Takagi.
\newblock Motivating multi-generational crowd workers in social-purpose work.
\newblock In {\em Proceedings of the 18th ACM Conference on Computer Supported
  Cooperative Work \& Social Computing}, pages 1813--1824. ACM, 2015.

\bibitem{kobayashi2013age}
Masatomo Kobayashi, Tatsuya Ishihara, Toshinari Itoko, Hironobu Takagi, and
  Chieko Asakawa.
\newblock Age-based task specialization for crowdsourced proofreading.
\newblock In {\em International Conference on Universal Access in
  Human-Computer Interaction}, pages 104--112. Springer, 2013.

\bibitem{mcmullin_generational_2007}
Julie~Ann McMullin, Tammy Duerden~Comeau, and Emily Jovic.
\newblock Generational affinities and discourses of difference: a case study of
  highly skilled information technology workers.
\newblock {\em The British Journal of Sociology}, 58(2):297--316, June 2007.

\bibitem{morrison_is_2013}
P.~Morrison and E.~Murphy-Hill.
\newblock Is programming knowledge related to age? {An} exploration of stack
  overflow.
\newblock In {\em 2013 10th {IEEE} {Working} {Conference} on {Mining}
  {Software} {Repositories} ({MSR})}, pages 69--72, May 2013.

\bibitem{nimrod2010seniors}
Galit Nimrod.
\newblock Seniors’ online communities: A quantitative content analysis.
\newblock {\em The Gerontologist}, 50(3):382--392, 2010.

\bibitem{nimrod2012online}
Galit Nimrod.
\newblock Online communities as a resource in older adults’ tourism.
\newblock {\em The Journal of Community Informatics}, 8(1), 2012.

\bibitem{nimrod2013probing}
Galit Nimrod.
\newblock Probing the audience of seniors’ online communities.
\newblock {\em The Journals of Gerontology Series B: Psychological Sciences and
  Social Sciences}, 68(5):773--782, 2013.

\bibitem{nimrod2014benefits}
Galit Nimrod.
\newblock The benefits of and constraints to participation in seniors’ online
  communities.
\newblock {\em Leisure Studies}, 33(3):247--266, 2014.

\bibitem{robat_history_2006}
C.~Robat.
\newblock The {History} of {Computing} {Project} ‘{TimeLine} {History} of
  {Computing}’.
\newblock Technical report, Accessed 02/03/2006) URL= http://www. thocp.
  net/timeline/timeline. htm, 2006.

\bibitem{slavicek_designing_2014}
T.~Slavicek, J.~Balata, and Z.~Mikovec.
\newblock Designing mobile phone interface for active seniors: {User} study in
  {Czech} {Republic}.
\newblock In {\em 2014 5th {IEEE} {Conference} on {Cognitive}
  {Infocommunications} ({CogInfoCom})}, pages 109--114, November 2014.

\end{thebibliography}
\bibliographystyle{plain}

\end{document}